\documentclass[11pt]{article}
\usepackage{amsmath}
\usepackage{amssymb}
\usepackage{graphicx}

\begin{document}

\setcounter{figure}{0}
\setcounter{table}{0}
\setcounter{footnote}{0}
\setcounter{equation}{0}

\noindent {\Large\bf Spherical Rectangular Equal-Area Grid (SREAG):\\ Some features\footnote{Presented at the
 Journ\'ees 2019 ``Astrometry, Earth rotation and Reference systems in the Gaia era'', Paris, France, 7-9 Oct 2019,
 https://syrte.obspm.fr/astro/journees2019/}}
 
\vspace*{0.5cm}
\noindent {\large\bf Zinovy Malkin}\\[0.1cm]
\noindent {Pulkovo Observatory, St. Petersburg, Russia, e-mail: malkin@gaoran.ru}\\

\vspace*{0.5cm}

\noindent {\large\bf ABSTRACT.}
A new method Spherical Rectangular Equal-Area Grid (SREAG) was proposed in Malkin (2019) for splitting spherical
surface into equal-area rectangular cells.
In this work, some more detailed features of SREAG are presented.
The maximum number of rings that can be achieved with SREAG for coding with 32-bit integer is $N_{ring}$=41068,
which corresponds to the finest resolution of $\sim$16$''$.
Computational precision of the SREAG is tested.
The worst level of precision is $7\cdot10^{-12}$ for large $N_{ring}$.
Simple expressions were derived to calculate the number of rings for the desired number of cells
and for the required resolution.

\vspace*{1cm}
\noindent {\large\bf 1. INTRODUCTION}
\smallskip

A new approach to pixelization of a spherical surface Spherical Rectangular Equal-Area Grid (SREAG) was proposed
in Malkin (2019).
It is aimed at constructing of a grid that best satisfies the following properties:

\begin{enumerate}
\itemsep=-0.7ex
\item it consists of rectangular cells with the boundaries oriented along the latitudinal and longitudinal circles;
\item it has uniform cell area over the sphere;
\item it has uniform width of the latitudinal rings;
\item it has near-square cells in the equatorial rings;
\item it allows simple realization of basic functions such as computation of the cell number given object position,
  and computation of the cell center coordinates given the cell number.
\end{enumerate} 
 
In this paper, some more details of the SREAG pixelization method are discussed in addition to Malkin (2019).

\vspace*{0.7cm}
\noindent {\large\bf 2. SREAG METHOD}
\smallskip

Let's briefly repeat the description of the SREAG pixelization method given in Malkin (2019). 
The basic parameter of this  method is the number of rings $N_{ring}$.
The sphere is first split into latitudinal $N_{ring}$ rings of constant width $dB=180^{\circ}/N_{ring}$.
Then each ring is split into several cells of equal size.
The longitudinal span of the cells in each ring is computed as $dL_i = dB \sec b_0^i$,
where $i$ is the ring number, and $b_0^i$ is the central latitude of the ring.
This provides near-square cells in the equatorial rings.
Then the number of the cells in each ring equal to $360 /dL_i$ is rounded to the nearest integer value.
This procedure results in the initial grid.
In fact, only the number of cells $N_{cell}$ and the number of the cells in each ring are used in the
final grid contraction.
Given $N_cell$ we can compute the area of each cell $A = 4\pi / N_{cell}$.

Then the latitudinal boundaries of the rings are to be adjusted as follows.
Let us start from the North pole.
Let $b^u$ be the upper (closer to the pole) boundary of the ring in the final grid,
and $b^l$ be the lower boundary.
Then, taking into account that the cell area is $A = dL * ( \sin b^u - \sin b^l )$, the simple loop will
allow to compute all the ring boundaries (Malkin, 2019):

\smallskip
\indent $b^u_1 = \pi/2$\\         
\indent do i=1,$N_{ring}/2$\\
\indent\indent $b^l_i = \arcsin( \sin b^u_i - A/dL_i )$\\
\indent\indent $b^u_{i+1} = b^l_i$\\
\indent end do
         
\begin{figure}
\centering
\includegraphics[clip,width=\hsize]{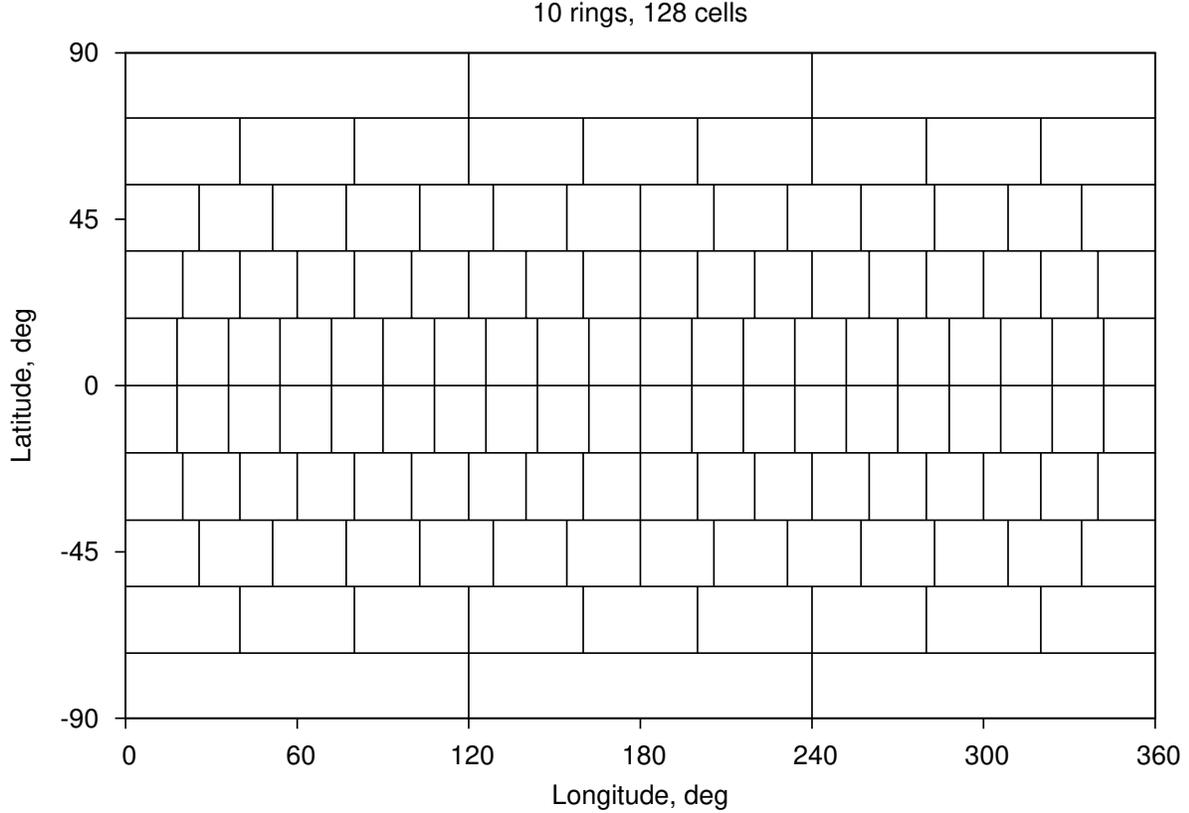}
\caption{Example: 10-ring SREAG grid.}
\label{fig:cells}
\end{figure}

\begin{figure}
\centering
\includegraphics[clip,width=\hsize]{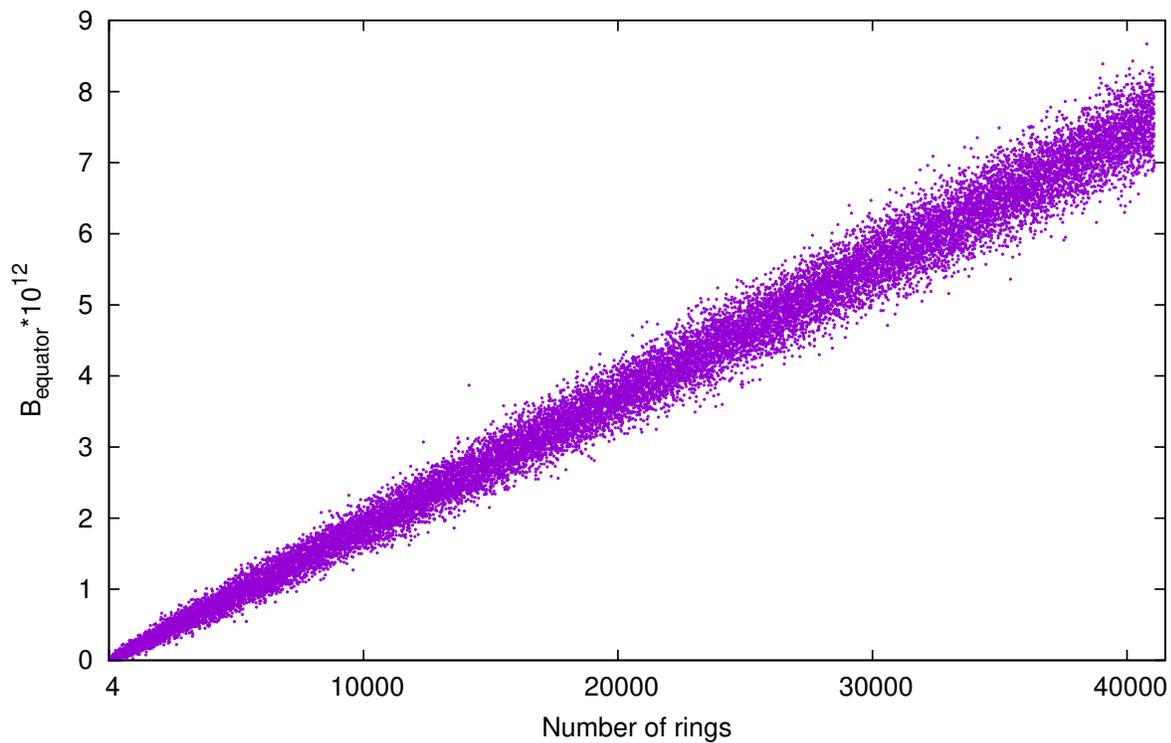}
\caption{Computational precision.}
\label{fig:precision}
\end{figure}

\medskip
The last value $b^l_{Nring/2}$ corresponds to the equator and therefore must be equal to zero, 
which verifies the correctness of the computation.
After that, the latitudinal boundaries for the rings in the South hemisphere are just copied from the North
hemisphere with negative sign.
Figure~\ref{fig:cells} presents an examples of grids constructed making use of the proposed method.
Figure~\ref{fig:precision} shows the precision of the computation that is defined by the deviation of the absolute
value of the computed equatorial latitude $b^l_{Nring/2}$ from zero.

\begin{figure}
\centering
\includegraphics[clip,width=\hsize]{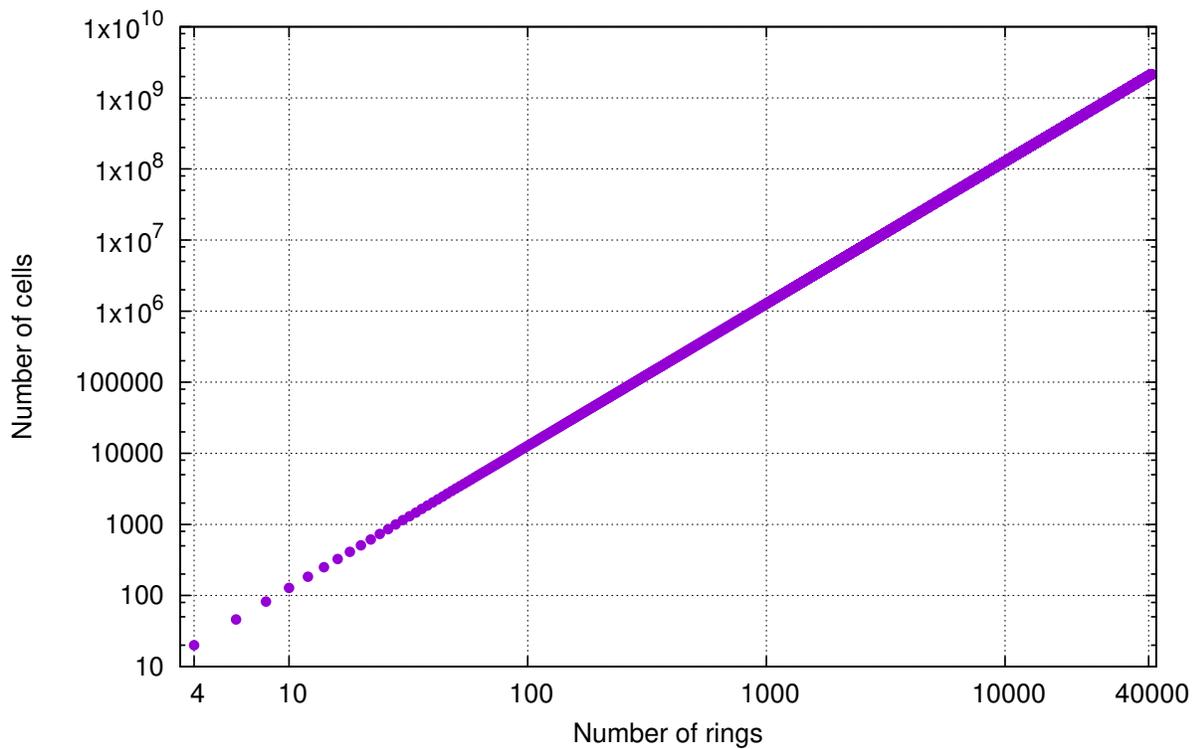}
\caption{The number of the cells in the grid as function of $N_{ring}$.}
\label{fig:nring_ncell}
\end{figure}

\begin{figure}
\centering
\includegraphics[clip,width=\hsize]{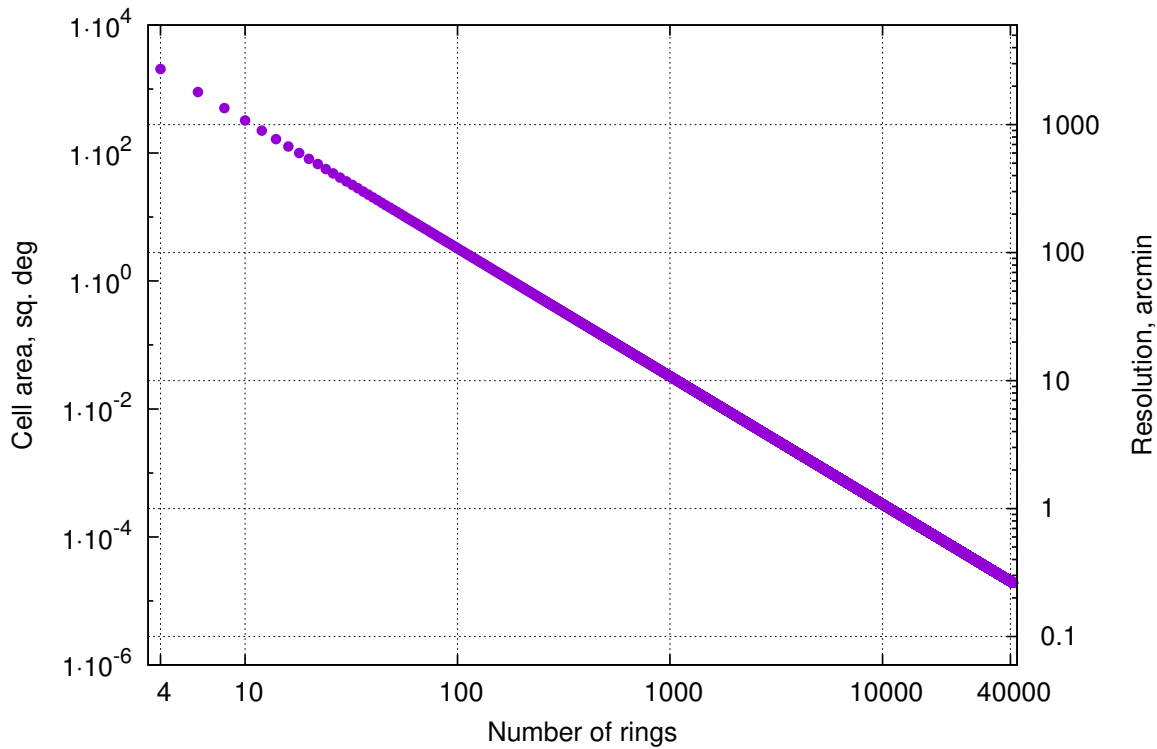}
\caption{Cell area and grid resolution.}
\label{fig:cell_area}
\end{figure}

The number of the cells in the grid depending on $N_{ring}$ is shown in Fig.~\ref{fig:nring_ncell}.
For 32-bit integer, maximum available $N_{ring}$ is 41068.
Therefore, the SREAG method provides detailed choice of the grid resolutions to satisfy a wide range
of user requirements.
For $N_{ring} = 4 \ldots 41068$ grid resolution varies from $\sim$45$^{\circ}$ to $\sim$16$''$
(Fig~\ref{fig:cell_area}).
Analysis of the literature showed that the resolution used in practice lies in the range 7.3$^{\circ}$ to 26$''$,
which is fully covered by the SREAG resolution range.
The latter can be extended using 64-bit integer.

If one starts with the desired $N_{cell}$, one can easily calculate the corresponding number of rings by
$N_{ring}^{\prime} = 0.886227 \, \sqrt{N_{cell}} \,$ with further rounding the result to the nearest even integer.

Another simple but accurate expression allows to approximate the grid resolution (in arcmin) as $10800/N_{ring}$
and thus obtain the required number of rings to provide the desired resolution.
Again, $N_{ring}$ must be even integer in the range [4:41068] for 32-bit compiler.

\clearpage

\vspace*{0.4cm}
\noindent {\large\bf 3. CONCLUSION}
\smallskip

The new method SREAG is developed for subdividing a spherical surface into equal-area cells.
The main features of the proposed approach are:
\begin{itemize}
\itemsep=-0.7ex
\item it provides an isolatitudinal rectangular grid cells with the latitude- and longitude-oriented boundaries with
  near-square cells in the equatorial rings;
\item it provides a strictly uniform cell area;
\item it provides a near-uniform ring width;
\item it provides a wide range of grid resolution with a possibility of detailed choice
  of desirable cell area;
\item the binned data is easy to visualize and interpret in terms of the longitude-latitude
  (right ascension-declinations) rectangular coordinate system, natural for astronomy and geodesy;
\item it is simple in realization and use.
\end{itemize}

Proposed approach to pixelization of a celestial or terrestrial spherical surface allows to construct a wide range
of grids for analysis of both large-scale and tiny-scale structure of data given on a sphere. 
The number of cells is theoretically unlimited and is constrained in practice only by the precision of machine
calculations.

The SREAG method can be hopefully useful for various practical applications in different research fields
in astronomy, geodesy, geophysics, geoinformatics, and numerical simulation. 
In particular, it can be used in further analyses of the celestial reference frame, for selection of uniformly
distributed reference sources in the next ICRF realizations, and for evaluation of the systematic errors of the source
position catalogs.

\clearpage

\noindent {\large\bf 4. SUPPORTING SOFTWARE}
\smallskip

Several Fortran routines to perform basic operations with SREAG are provided at

\verb"http://www.gaoran.ru/english/as/ac_vlbi/#SREAG" . They include:

\begin{tabular}{ll}
GRIDPAR.FOR & Compute parameters of the grid for a given number of rings. \\
CELLPAR.FOR & Compute the cell parameters for a given cell number. \\
POS2CN2.FOR & Compute the cell number for a given point position. \\
CN2POS2.FOR & Compute the cell center coordinates for a given cell number. \\
NR2NC.FOR   & Compute the number of cells for a given number of rings. \\
NC2NR.FOR   & Compute the nearest number of rings for a given number of cells. \\
\end{tabular}


\vspace*{0.7cm}
\noindent {\large\bf 5. REFERENCES}

{

\leftskip=5mm
\parindent=-5mm
\smallskip

Malkin, Z. (2019) A new equal-area isolatitudinal grid on a spherical surface, AJ, Vol.~158, id.~158,
doi:~10.3847/1538-3881/ab3a44.

}

\end{document}